\documentclass{elsart}
\usepackage{amssymb,psfig,subfigure,graphicx}

\def\i{{\rm i}\,}
\newcommand{\dsum}{\displaystyle \sum\limits}
\newcommand{\eps}{\varepsilon}
\newcommand{\fat}[1]{\mbox{\boldmath $ #1 $\unboldmath}}
\newcommand{\beq}{\begin{equation}}
\newcommand{\beqa}{\begin{eqnarray}}
\newcommand{\eeq}{\end{equation}}
\newcommand{\eeqa}{\end{eqnarray}}
\newcommand{\bigfrac}[2]{\mbox {${\displaystyle \frac{ #1 }{ #2 }}$}}
\newcommand{\ket}[1]{\left | #1 \right\rangle}
\def\dwn{\downarrow}
\def\up{\uparrow}
\def\d{\dagger}
\newcommand{\comm}[2]{\left[ #1, #2 \right]}
\renewcommand{\t}[1]{\tilde{#1}}

\begin{document}
\begin{frontmatter}
\title{Integrable  models for  confined fermions: applications 
to metallic grains}
\author[DMFCI,INFM]{Luigi Amico},
\author[DMFCI,INFM]{Antonio Di Lorenzo}, and
\author[DMFCI,INFM]{Andreas Osterloh}
\address[DMFCI]{Dipartimento di Metodologie Fisiche e Chimiche (DMFCI), 
	Universit\`a di Catania, viale A. Doria 6, I-95125 Catania, Italy}
\address[INFM]{Istituto Nazionale per la Fisica della Materia, Unit\`a di Catania, 
NEST-INFM}

\begin{abstract}
We study integrable models for electrons in metals when 
the single particle spectrum is discrete. 
The electron-electron interactions are  BCS-like pairing,  
Coulomb repulsion, and  spin exchange coupling.
These couplings are, in general, nonuniform in the sense that 
they depend on the levels occupied by the interacting electrons.
By using the realization of  spin $1/2$-operators in terms 
of electrons the models describe spin $1/2$ models 
with nonuniform long range interactions and external 
magnetic field. 
The integrability and the exact solution arise 
since the model Hamiltonians can be  constructed in terms 
of Gaudin models. 
Uniform pairing and the resulting orthodox model 
correspond to an isotropic limit of the Gaudin Hamiltonians.
We discuss possible applications of this model 
to a single grain 
and to a system of few interacting grains. 
\end{abstract}
{\it PACS} N. 74.20.Fk, 02.30.Ik
\end{frontmatter}

\section{Introduction}
Recent advances in nanophysics allowed experimental investigations on 
isolated quantum dots and metallic nanoparticles\cite{RBT}. 
Because of their small size, such systems have a discrete spectrum, 
which can be resolved through I-V characteristic measurements. 
Besides potential technological applications, 
spectral analysis provides insight into
the nature of the interactions in metals, 
comparable to the spectroscopy-based knowledge about the interactions 
in atoms and molecules\cite{REPORT}.
This motivates the broad recent interest in various physical contexts: 
the spin orbit interaction was studied in 
mesoscopic $Al$-$Au$ systems\cite{DAVIDOVIC};
various ferromagnetic properties were analyzed in experiments on 
$C\!o$ grains\cite{GUERON};
the experiments on $Al$ samples opened up debates on the crossover 
behavior of superconducting fluctuations, when the size of the 
sample is progressively reduced. 
The peculiarity of small grains and quantum dots is
their small capacitance, which fixes the number of electrons in the system. 
In contrast, the characterization of many physical phenomena 
like e.g. phase transitions relies on quantities defined for infinite systems 
in the grand-canonical description. 
Therefore, the fingerprints of these phenomena, i.e. their 
characterizing physical quantities, must be identified at first\cite{VONDELFT}.
Then, techniques suitable for systems in the canonical ensemble 
must be applied.
Finally, for such small systems, approximations are difficult to control 
because of the presence of strong finite-size fluctuations and then,
exact results are extremely valuable.

Many efforts have been devoted to the construction of theoretical models 
describing the physics of small metallic grains and 
quantum dots~\cite{GLAZMAN,ALTSHULER,STONE,ALHASSID}. 
It has been argued that, due to the underlying disorder, 
the leading part of the electron-electron interaction 
reduces the Hamiltonian to a simple ``universal'' form
(i.e. independent of the geometry of the system 
and the realization of the disorder), 
containing only the electrostatic interaction $E_c$, 
a BCS-like coupling $g$, and a spin-spin coupling $J$
\begin{equation}
H_U = \sum_{i\sigma} \varepsilon_{i\sigma} n_{i\sigma} + E_{c} N^2 
- J \vec{S}^2 - g K^+ K^-, 
\label{uniH}
\end{equation}
where the $\varepsilon_{i}$ form the effective single-particle spectrum, 
$n_{i\sigma}$ is the electron number operator, 
$\vec{S}$ and $N$ are the total spin and number of electrons respectively. 
$K^-=\sum_i c^{}_{i\dwn} c^{}_{i\up}$, $K^+={K^-}^\dagger$ are the 
pair annihilation and creation operators. 
We point out that the couplings in (\ref{uniH}) are independent of
the involved single-particle levels; in this sense, they are uniform.
Non-universal corrections to Hamiltonian (\ref{uniH}) are of relative order 
$\delta/E_{Th}$, $E_{Th}$ being the Thouless 
energy and $\delta$ the average level spacing.\cite{GLAZMAN,ALTSHULER}

Hamiltonian (\ref{uniH}) has been used to attack a variety of physical
phenomena\cite{PROCGRAINS}: equilibrium transport near degeneracy have been 
thoroughly investigated~\cite{GLAZMAN}; recently,
Kurland {\it et al.} have shown that the 
paramagnetic phase is favored by the presence of disorder in the 
single-particle level spacing of the $\varepsilon_i$~\cite{ALTSHULER}.
\\
For $g=0$, the Hamiltonian (\ref{uniH}) describes a normal 
metallic grain with spin-spin interaction. Its eigenstates are 
certain linear combinations of 
Fock states (eigenstates of $n_{i\sigma}$), 
such to be eigenstates of $\vec{S}^2$ as well; 
they are obtained applying the angular momentum lowering operator
to the highest weight state.
\\
For $g\neq 0$ 
the exact solution of $H_{U}$ was found by 
Richardson and Sherman (RS)~\cite{RICHARDSON} 
using techniques close in spirit to the coordinate Bethe Ansatz (BA). 
By expressing $H_{U}$ in terms of commuting integrals of motion 
in Refs.~\cite{CRS,ADO} (see also Ref.~\cite{AMICO}), 
it was demonstrated that this Hamiltonian is also integrable.
In fact, the exact BA eigenstates of $H_{U}$ have been 
successfully employed in the recent literature
\cite{VONDELFT,SIERRADUKELSKY}.

Non-equilibrium transport experiments~\cite{AGAM} 
show that the corrections ${\mathcal O}(\delta/E_{Th})$ to 
Hamiltonian (\ref{uniH}) become very important for ultra-small samples. 
The presence of resonance clusters in the tunneling conductivity 
through Al grains are experimental evidence 
of fluctuations in the electrostatic interaction; in quantum dots 
the universal Hamiltonian (or Constant Interaction model) predicts a 
``bimodal'' distribution of the tunneling-peak spacing, 
while a distribution corresponding to a Gaussian Orthogonal Ensemble (GOE) 
is observed~\cite{GLAZMAN}. 
Further, it has been argued that fluctuations in the spin pairing
should influence the threshold for the mesoscopic Stoner 
instability, which could be observed in future experiments\cite{STONE}. 
The presence of fluctuations leads to {\it nonuniform couplings} 
in the Hamiltonian, i.e. couplings whose strength depends on the 
two interacting levels. 
Not all nonuniform couplings yield an exactly solvable Hamiltonian;
of course, demanding solvability results in restrictions 
on the admitted form of the interactions.
\\
The aim of the present work is to explore the possible forms of the 
couplings compatible with the exact solvability of the corresponding models. 
We could not find\cite{TRIVIALEXTENSION} solvable generalizations
of $H_{U}$ starting from a quadratic bosonic model, as done by 
RS. The strategy we adopt, instead, consists in 
generalizing the procedure of Ref.~\cite{CRS}, 
namely constructing the Hamiltonian of the system 
in terms of Gaudin Hamiltonians.
By means of the integrability and the exact solvability of the latter
the exact solution of the model is obtained.
For uniform couplings the integrals of motion are tightly related to the 
isotropic Gaudin Hamiltonians; for nonuniform couplings to the 
anisotropic Gaudin Hamiltonians.
The models arising from this procedure include certain nonuniform 
interactions, which could account for corrections to the universal Hamiltonian
due to e.g. disorder fluctuations. Namely: 
\beq
H = \sum_{i\sigma} \eps_{i\sigma} n_{i\sigma} 
	+ \sum_{ij}
	U_{ij} \; n_{i} n_{j} - \sum_{ij} g_{ij} \; c^\d_{i\up} c^\d_{i\dwn} 
	c^{}_{j\dwn} c^{}_{j\up} -\sum_{ij} J_{ij} \vec{S}_i \cdot \vec{S_j}   ,
\eeq
where $U_{ij}$, $g_{ij}$ and $J_{ij}$ are the nonuniform generalizations of $E_c$, $g$ and $J$ 
in Eq.(\ref{uniH}), respectively. 
This class of models can also be interpreted as describing  
systems of several coupled grains. Then, 
the coupling constants necessarily have to differ significantly when 
describing intra- and inter-grain interaction respectively. 
Other possible applications may be found in nuclear physics, 
where the nonuniformity of the couplings is notable. 

This article is structured as follows: 
In section \ref{model}, 
we review how to derive the form of the Hamiltonian 
describing small grains, including the effects of disorder. 
Then we will rewrite the Hamiltonian in terms of spin and charge 
realizations of su(2). This will shed light on its structure, and 
lead to useful simplifications in the diagonalization procedure. 
In section \ref{uniform}, 
we sketch the solution of the universal part of the Hamiltonian and present
the isotropic Gaudin Hamiltonians and their connection with the integrals of 
motion. 
In section \ref{nonuniform}, 
we construct integrable Hamiltonians by means of {\em anisotropic} 
Gaudin Hamiltonians. This class of Hamiltonians goes beyond the 
universal model with constant couplings. In general it is
quadrilinear in the pseudo-spin operators. The tri- and quadrilinear terms
can be eliminated by a suitable choice of parameters.
In section \ref{discuss}, 
we discuss regimes of the interactions for exemplary choices of the
parameters.
Section~\ref{concls} is dedicated to concluding remarks.
In the appendix, we review some exact results obtained by Gaudin.

\section{The model Hamiltonian}\label{model}

In 2nd quantization, the Hamiltonian for $N$ particles in 
a confining potential $\hat{V}_1=\sum_a V(\hat{\bf r}_a)$ and a
(spin-independent) electron-electron interaction 
$\hat{V}_2=\frac{1}{2} 
\sum_{a\neq b} v(\hat{\bf r}_a-\hat{\bf r}_b)$, $a,b\in\{1,\dots,N\} $ is
\begin{eqnarray}
H&=& \sum_a\frac{{\hat{\bf p}_a}^2}{2m} + \hat{V}_1 + \hat{V}_2 = 
\sum_{i\sigma} \t{\varepsilon}_{i\sigma} n_{i\sigma} + 
\sum_{ijkl} \sum_{\sigma\tau} M^{ij}_{kl} c^\d_{i\sigma} c^\d_{j\tau}c^{}_{l\tau} c^{}_{k\sigma}, \label{genham}\\
M^{ij}_{kl}&=&
\frac{1}{2} \int d{\bf r} d{\bf r'} \phi_{i}^*({\bf r})\phi_{j}^*({\bf r'}) v({\bf r - r'}) 
\phi^{}_{l}({\bf r'})\phi^{}_{k}({\bf r}), \nonumber
\end{eqnarray}
where $\t{\varepsilon}_{i\sigma}$ and $\phi^{}_i({\bf r}) \chi_i(\sigma)$, 
$i \in  \{1, ... , \Omega\}:= I $
are the eigenvalues and eigenfunctions of the single particle Hamiltonian 
($\chi(\sigma)$ is the spinor basis); 
$c^\d_{i\sigma}$, $c^{}_{i\sigma}$ are electronic 
creation and annihilation operators.
The disorder average reduces the leading part $\bar{M}$ of the 
matrix elements $M^{ij}_{kl}$ to be ``uniform-diagonal''; 
i.e. only elements corresponding to pairwise equal indices 
are non-zero and constant:
$\bar{M}^{ij}_{ij}:=\alpha$, $\bar{M}^{ij}_{ji}:=\beta$ 
and $\bar{M}^{ii}_{jj}:=\gamma$ for $i\neq j$, $\bar{M}^{ii}_{ii}:=\omega$.
We can rewrite the interaction matrix as  
\beqa
M^{ij}_{kl}&=&
\delta_{ik} \delta_{jl}(1-\delta_{ij}) \alpha + 
\delta_{il} \delta_{jk}(1-\delta_{ij}) \beta +
\delta_{ij} \delta_{kl}(1-\delta_{ik}) \gamma \label{2body}\\
&&+ \delta_{ij}\delta_{jk}\delta_{kl} \omega + 
{\delta M}^{ij}_{kl};\nonumber 
\eeqa
the (non-universal) fluctuations ${\delta M}^{ij}_{kl}$
are such that
$\overline{{{\delta M}^{ij}_{kl}{\delta M}^{ij}_{kl}}}= {\mathcal O}(\delta^2/E_{Th}^2)$, i.e.   
the corrections are of relative order $\delta/E_{Th}$. 
Inserting expression (\ref{2body}) in the Hamiltonian (\ref{genham}), 
we obtain 
$H=H_U + \delta H$, where
\begin{eqnarray}
H_U &=&\sum_{i\sigma} (\t{\varepsilon}_{i\sigma} + 
\delta\!M_{ii}^{ii}) n_{i\sigma} +
\sum_{i\neq j} \sum_{\sigma\tau} \Biggl[ \alpha n_{i\sigma} n_{j\tau} + 
\beta c^\d_{i\sigma} c^\d_{j\tau} c^{}_{i\tau} c^{}_{j\sigma} + 
\gamma c^\d_{i\sigma} c^\d_{i\tau}c^{}_{j\tau} c^{}_{j\sigma} \Biggr] 
\nonumber \\
& &+ \sum_i \sum_{\sigma\tau} 
\omega~c^\d_{i\sigma} c^\d_{i\tau}c^{}_{i\tau} c^{}_{i\sigma} \nonumber \\
 &=& \sum_{i\sigma} \varepsilon_{i\sigma} n_{i\sigma} + 
 \sum_{ij} \Biggl[ 
\alpha n_i n_j -  
\beta \sum_\sigma 
(c^{\d}_{i\sigma} c^{ }_{i\bar{\sigma}} c^{\d}_{j\bar{\sigma}} 
c^{ }_{j{\sigma}} + n_{i\sigma} n_{j\sigma}) 
+ 2 \gamma c^\d_{i\up} c^\d_{i\dwn} c^{}_{j\dwn} c^{}_{j\up} \Biggr] \nonumber \\
&&+ 2\left( \omega-\alpha-\beta-\gamma\right) N_{pairs}  \nonumber \\ 
\delta H &=& 
\sum_{ij} \left[
{\delta\!M}^{ij}_{ij} n_{i} n_{j} -
{\delta\!M}^{ij}_{ji} \sum_\sigma 
(c^{\d}_{i\sigma} c^{ }_{i\bar{\sigma}} c^{\d}_{j\bar{\sigma}} c^{ }_{j{\sigma}} + 
n_{i\sigma} n_{j\sigma})
+ 2~{\delta\!M}^{ii}_{jj} c^\d_{i\up} c^\d_{i\dwn}c^{}_{j\dwn} c^{}_{j\up} 
\right] \nonumber \\
&&- \sum_i 4~\delta\!M^{ii}_{ii} n_{i\up} n_{i\dwn} \nonumber\\
&&+ \sum_{ijkl} \sum_{\sigma\tau} 
(1-\delta_{ij}\delta_{kl})(1-\delta_{ik}\delta_{jl})(1-\delta_{il}\delta_{jk})
M_{kl}^{ij} c^\d_{i\sigma} c^\d_{j\tau}c^{}_{l\tau} c^{}_{k\sigma}, \nonumber 
\end{eqnarray}
where $n_i=\sum_\sigma n_{i\sigma}$ is the number of electrons in the $i$-th \emph{level-pair}, 
$\bar{\sigma}=-\sigma$, 
$N_{pairs}=\sum_i n_{i\up} n_{i\dwn}$ is the number of pairs, and
$\varepsilon_{i\sigma}=\t{\varepsilon}_{i\sigma} - \alpha + 2\beta + \delta\!M^{ii}_{ii}$.
Introducing the spin operators
\begin{equation}
S_j^+ = c^\d_{j\up} c^{}_{j\dwn} \quad, \quad S_j^- = \left(S_j^+\right)^\d \quad, \quad
S_j^z = \frac{1}{2} \left(n_{j\up} - n_{j\dwn}\right), 
\end{equation}
we have $n_{i\up}=\frac{1}{2} n_{i} + S^z_i$, $n_{i\dwn}=\frac{1}{2} n_{i} - S^z_i$, 
and then: 
\begin{eqnarray}
H_U &=&\sum_{i\sigma} {\varepsilon}_{i\sigma} n_{i\sigma} +
\sum_{ij} \left[(\alpha-\frac{\beta}{2}) n_i n_j - 
\beta \vec{S}_i \!\cdot\! \vec{S_j} 
+ 2\gamma c^\d_{i\up} c^\d_{i\dwn}c^{}_{j\dwn} c^{}_{j\up} \right]
\label{uniH2} \\ 
&& + 2\left( \omega-\alpha-\beta-\gamma\right) N_{pairs}  \nonumber\\
\delta H &=& \sum_{ij} \left[\left({\delta\!M}^{ij}_{ij}
- \frac{1}{2} {\delta\!M}^{ij}_{ji}\right) n_{i} n_{j} -
{\delta\!M}^{ij}_{ji} \vec{S}_i \!\cdot\! \vec{S_j} 
+ 2~{\delta\!M}^{ii}_{jj} c^\d_{i\up} c^\d_{i\dwn}c^{}_{j\dwn} c^{}_{j\up} 
\right] \nonumber \\ 
&&- \sum_i 4~\delta\!M^{ii}_{ii} n_{i\up} n_{i\dwn} \label{disH2} \\
&& + \sum_{ijkl} \sum_{\sigma\tau} 
(1-\delta_{ij}\delta_{kl})(1-\delta_{ik}\delta_{jl})(1-\delta_{il}\delta_{jk})
M_{kl}^{ij} c^\d_{i\sigma} c^\d_{j\tau}c^{}_{l\tau} c^{}_{k\sigma}. \nonumber 
\end{eqnarray}
From $H_U$ we recover, up to a constant of motion, the universal Hamiltonian (\ref{uniH}), 
defining $E_{c}\equiv \alpha - \beta/2$, $J\equiv\beta$ 
and $g\equiv-2 \gamma$. 
This constant of motion is relevant to 
determine the ground state properties. 
\\
The Hamiltonian studied within the present paper is
the sum of the uniform part (\ref{uniH2}) 
and the first line of the nonuniform correction (\ref{disH2}):
\beqa
H &=& \sum_i 2\zeta_i S^z_i 
	-\sum_{ij} J_{ij} \vec{S}_i \cdot \vec{S_j} \nonumber \\
&&+ 
	\sum_i \xi_i n_{i} + \sum_{ij}
	U_{ij} \; n_{i} n_{j} - \sum_{ij} g_{ij} \; c^\d_{i\up} c^\d_{i\dwn} 
	c^{}_{j\dwn} c^{}_{j\up} ,
\label{ourH} 
\eeqa
where we defined $2 \zeta_i = \varepsilon_{i\up}- \varepsilon_{i\dwn}$, 
$2\xi_i= \varepsilon_{i\up} + \varepsilon_{i\dwn}$.
It can be written in a more perspicuous way introducing the operators
\beq
K_j^+ = c^\d_{j\up} c^\d_{j\dwn},\quad \quad K_j^- = 
\left(K_j^+\right)^\d , \qquad 
K_j^z = \frac{1}{2} \left( n_{j\up} + n_{j\dwn} - 1 \right) ,
\eeq
which are generators of the charge $su(2)$:
$\comm{K_i^\pm}{K_j^z}=\pm \delta_{ij} K_i^\pm$, 
$\comm{K_i^+}{K_j^-}=2 \delta_{ij} K_i^z$.
The charge $su(2)$ is orthogonal to the spin $su(2)$, 
i.e. their generators have the property $\comm{S_j^a}{K_k^b} = 0$. 
In terms of the spin and charge $su(2)$ operators, 
the Hamiltonian ({\ref{ourH}}) can be finally written as 
\begin{eqnarray}
H &=& H_S + H_K + E_0 \label{totalHam} \\
H_S &=& \sum_i 2 \zeta_i S_i^z - \sum_{ij} J_{ij}\: \vec{S}_i \cdot \vec{S}_j \quad , 
\label{spinh} \\ 
H_K &=& \sum_i 2 \eta_i K_i^z + 4 \sum_{ij} U_{ij} K_i^z K_j^z - \frac{1}{2} \sum_{ij} 
 g_{ij} (K^+_i K^-_j + K_i^- K_j^+)  , 
\label{chargeh} 
\end{eqnarray}
where $2\eta_i=2\xi_i + 4 \sum_j U_{ij} - g_{ii}$ and
$E_{0}=\sum_i \xi_i + \sum_{ij} U_{ij}$. \\
In what follows, we refer to $H_S$ as \emph{spin} Hamiltonian, and to $H_K$ as \emph{charge} 
Hamiltonian. They belong to the class of integrable models we find in section \ref{nonuniform}.
\\ 
We note that $[H, S^z] = [H, K^z] = 0$. Hence we can diagonalize the Hamiltonian in the 
subspace of fixed number of pairs $N_K$ and up spins $N_S$. 
Because of the orthogonality of the spin and charge $su(2)$, 
it follows that $[H_K,H_S]=0$ and in particular 
$[H_S,K^\pm_i]=[H_K,S^\pm_i]=0$. Thus  
$H_S$ and $H_K$ are degenerate respect to the action of 
$K^\pm_i$ and $S^\pm_i$ respectively.
Therefore, the Hilbert space ${\mathcal H}$ can be split into 
${\mathcal H}_S$ and ${\mathcal H}_K$ created by the action of  
$N_S$ $S^+_i$ and $N_K$
$K^+_i$ on the  vacua $\ket{0}_S$ and $\ket{0}_K$ respectively. 
The  {\it spin vacuum}   $\ket{0}_S=\ket{\downarrow,\dots,\downarrow}$ 
is the state in which all electrons of the singly occupied level-pairs 
have  spin down; 
the {\em charge vacuum} $\ket{0}_K$ is the state in ${\mathcal H}_K$ 
with all level-pairs empty.
Namely, ${\mathcal H}_K$ consists of both 
the doubly occupied and empty level-pairs which we label with 
the index set $I_K$ ($|I_K|=\Omega_K$), and ${\mathcal H}_S$ consists 
of the singly occupied levels-pairs, which we label by the index set 
$I_S=I\setminus I_K$ ($|I_S|=\Omega_S$).
This factorization of the Hilbert space 
was called the ``blocking of singly occupied levels'' when considering 
the charge Hamiltonian only\cite{RICHARDSON}, and it 
reduces the problem of finding the spectrum of $H$
to the diagonalization of $H_S$ and $H_K$ in 
${\mathcal H}_S$ and ${\mathcal H}_K$, respectively.

\section{Uniform couplings}\label{uniform}

In this section, we recast the  known results\cite{VONDELFT,RICHARDSON,CRS} 
for the universal Hamiltonian in a uniform 
magnetic field  in the present frame. 
In this case, having decomposed the universal model (\ref{uniH}) 
as in formula (\ref{totalHam}),
the spin and charge Hamiltonians are
\begin{eqnarray}
H_S&=& 2 \zeta S^z - J {\vec{S}}^2, \nonumber \\
H_K&=&\sum_i 2 \eta_i K_i^z - \frac{g}{2} \sum_{ij} (K_i^+ K_j^- + K_i^- K_j^+) 
+ 4 E_{c} \left(\sum_{i} K_i^z\right)^2 , 
\end{eqnarray}
and $E_0=\sum_i \xi_i + \Omega_K^2 E_c$, 
while the parameters are $\zeta =\mu_B B$ and $2\eta_i=2\xi_i + 4 E_{c} \Omega_K - g$.
The eigenstates 
$\ket{\Psi}_{N_S}=(S^+)^{N_S}\ket{0}_S$ 
of $H_S$ are common eigenstates of 
the total spin $\vec{S}^2$ and $S^z$ 
with eigenvalues $E_{N_S} = 2 \mu_B B (N_S - \Omega_S/2) - J S (S+1)$.

$H_K$ can be diagonalized through coordinate~\cite{RICHARDSON}
or algebraic~\cite{BRAUN} BA. 
Due to its integrability, it suffices to diagonalize its
integrals of motion $\tau_i$.
Indeed, $H_K$ can be written as 
\begin{eqnarray}
H_K &=& \sum_{i} 2\eta_i \; \tau_i 
+ (4E_c + g) \sum_{ij} \tau_i\; \tau_j + const. ,
\label{bcs} \\
\tau_i ({\bf K}) &=& K_i^z - 
g \sum_{j\neq i}\frac{\vec{K}_i\cdot\vec{K}_j}{\eta_i-\eta_j} 
:= K_i^z + \Xi_i.
\label{isointmot}
\end{eqnarray}
where $\Xi_i$ are the isotropic Gaudin Hamiltonians (see also the appendix).
With a straightforward generalization of 
Gaudin's results\cite{GAUDIN,SKLYANIN-GAUDIN}, 
one recovers eigenstates and eigenvalues of $\tau_i$, 
and from these the eigenvalues $E_{N_K}$ of $H_K$
\begin{eqnarray}
\ket{\Psi_{N_K}}&=&\prod_\alpha \sum_i 
\frac{K_i^+}{E_\alpha - 2\xi_i} \ket{0}_K \nonumber \\
E_{N_K}&=&\sum_\alpha E_\alpha + E_c (2 N_K-\Omega_K)^2\,. 
\label{eigenbcs}
\end{eqnarray}
The quasi-energies $E_\alpha$ are solutions of the
RS equations
\begin{equation} 
-\frac{1}{g} + \sum_{j} \frac{1}{2 \xi_j - E_\alpha} 
= 2 {\sum_{\beta}}' \frac{1}{E_\beta - E_\alpha} \qquad \alpha= 1\dots N_K.
\label{richardson}
\end{equation}
Here and in the rest of the paper, a primed sum 
means that coincidences of indices  are avoided, e.g.
$\sum_\alpha 'f(\alpha,\beta):=
\sum_{\alpha \atop \alpha\neq\beta} f(\alpha,\beta)$.

\section{nonuniform couplings}\label{nonuniform}

In this section we consider the following class of models, which includes 
Hamiltonian~(\ref{totalHam}), 
\beqa
\t{H}_K &=& \sum_i 2 \eta_i K_i^z - \frac{1}{2}
\sum_{ij} g_{ij} (K^+_i K^-_j + K_i^- K_j^+) + 
4 \sum_{ij} U_{ij} K_i^z K_j^z \label{genHamK}\\
\t{H}_S &=& \sum_i 2 \zeta_i S_i^z - \frac{1}{2}\sum_{ij} 
J_{ij} (S_i^+ S_j^- + S_i^- S_j^+) - 
\sum_{ij} J^z_{ij} S_i^z S_j^z  , \label{genHamS}
\eeqa
We prove the integrability of this class 
for a particular choice of the coupling constants 
(Eq.~(\ref{couplings}) below), and find its exact solution. 
The proof of integrability is constructive and performed for the
spin and charge Hamiltonian separately 
(see the discussion at the end of section \ref{model}). 
We summarize the results of subsections A and B.

We find that $H_K$ and $H_S$ are integrable if the couplings are 
\beq\label{intcouplings1st}
\left.
\begin{array}{rcl}
\label{couplings}
g_{ij} &=& - \bigfrac{p u(\eta_{i}-\eta_{j})}{
	\sinh[p (u_{i}-u_{j})]} \;, \\
&& \\
4 U_{ij} &=& A +  p u (\eta_{i} - \eta_{j}) \coth[p (u_{i}-u_{j})] \; , \\
&& \\
J_{ij} &=& - \bigfrac{ q v{\left(\zeta_{i}-\zeta_{j}\right)}}
{\sinh [q (v_{i}-v_{j})]} , \\
&& \\
J^z_{ij} &=& - q v (\zeta_{i} - \zeta_{j}) \coth{[q (v_{i}-v_{j})]},
\end{array}
\right\} \quad \mbox{for }i\neq j 
\eeq
where the parameters $A$, $u$, $v$, $u_i$, $\eta_i$ and $v_i$ are 
arbitrary real numbers, while $p$ and $q$ are imaginary or real. 
A quantitative discussion of the couplings for an 
exemplary and physically reasonable choice of parameters, 
is done in section \ref{discuss}. We remark that no analog to $A$ occurs 
for the spin term. The reason is that we can express $H_S$
even without the bilinear term in the integrals of motion, and that
$\sum_{i,j} \tau_i({\bf S}) \tau_j({\bf S}) = (S^z)^2$ is an integral of motion due
to a symmetry (the $S^z$-symmetry of the Hamiltonian) which is not
related with integrability. For $H_K$, instead, this constant is needed 
to keep the electron-electron Coulomb interaction repulsive.
\\ 
For every partition of the set $I$ in distinct subsets $I_S$, $I_K$, 
we obtain the eigenstates
\beqa\label{eigenstate1st}
\ket{\Psi} &=& \prod_{\alpha=1}^{N_K}\sum_{i\in I_K} 
\frac{p~K_i^+}{\sinh{[p(\omega_\alpha - u_i)]}}
\prod_{\beta=1}^{N_S}\sum_{j\in I_S} 
\frac{q~S_j^+}{\sinh{[q(\nu_\beta - v_j)]}} 
\ket{0}_K \otimes \ket{0}_S
\eeqa
whose energy is $E = E_{N_K} + E_{N_S}+E_0$, 
where $E_{N_K}$ and $E_{N_S}$ are given by
\beq
E_{N_K} = p u \sum_{j\alpha} \eta_{j} \coth{[p ( \omega_\alpha - u_j )]}
- A N_K (\Omega_K - N_K) - E_K
\label{eigencharge1st}
\eeq
\begin{equation}
E_{N_S} = p v \sum_{j\alpha} \zeta_{j} \coth{[p ( \nu_\alpha - v_j )]}
- \frac{1}{4}\sum_{ij} J^z_{ij} -\frac{1}{2} \sum_i J^{ }_{ii}\, .
\label{spineigen1stXXZ}
\end{equation}, 
where $E_K = \sum_i \left( \xi_i + \sum_j U_{ij}         
+ 2 \sum_{j\in I_S} U_{ij}\right)$.
From here on, latin indices, when referring to the charge 
(spin) Hamiltonian, range in $I_K$ ($I_S$), and greek indices from 1 to $N_K$ 
($N_S$). 
The parameters $\omega_\alpha$, $\nu_\beta$ 
must fulfill the following set of equations
\beqa
\frac{2}{u} + \sum_j p\;\coth{p[(u_j-\omega_\alpha)]} 
&=& 2{\sum_{\beta}}' 
p\; \coth{[p ( \omega_\beta - \omega_\alpha )]} ,
\label{generalized-richardson1st} \\
\frac{2}{v} + \sum_j q\;\coth{q[(v_j-\nu_\alpha)]} 
&=& 2{\sum_{\beta}}' 
q\; \coth{[q ( \nu_\beta - \nu_\alpha )]} \, .
\label{generalized-richardson1stspin}
\eeqa
For XXX-type coupling, present in the spin Hamiltonian (\ref{spinh}), 
that is $J^z\equiv J$, the coupling has the form
\beq
J_{ij}^{} = - v \bigfrac{\zeta_{i}-\zeta_{j}}{
   	v_{i}-v_{j}}\quad \mbox{for } i\neq j 
\label{spineigen1st}
\eeq
the eigenstates are then
\beq
\ket{\Phi^{xxx}_{N_S}}=\prod_{\beta}\sum_{j} 
\frac{S_j^+}{\nu_\beta - v_j} \ket{0}_S
\label{eigenstate}
\eeq
with eigenenergy
\beq
E_{N_S} = \sum_{j\alpha} \frac{v \zeta_i}{\nu_\alpha - v_i} 
-\sum_j \zeta_j -\frac{1}{4} \sum_{ij} J_{ij} - 
\frac{1}{2} \sum_i J^{ }_{ii}\,.
\label{nonunispineigen1st}
\eeq
The quantities $\nu_\beta$ must fulfill the RS equations
\begin{equation} 
\frac{2}{v} + \sum_{j} \frac{1}{v_j - \nu_\alpha} 
= 2 {\sum_{\beta}}' \frac{1}{\nu_\beta - \nu_\alpha}.
\label{modigaudin1st}
\end{equation}
Finally, for uniform Zeeman splitting $\zeta_j\equiv \mu_B B$,
the eigenstates are still given by (\ref{eigenstate}), but
the Gaudin equations
\begin{equation}	
\sum_{j} \frac{1}{v_j - \nu_\alpha} 
= 2 {\sum_{\beta}}' \frac{1}{\nu_\beta - \nu_\alpha}.
\label{gaudineq1st}
\end{equation}
have to be fulfilled instead of (\ref{modigaudin1st}).
The eigenenergy $E_{N_S}$ is then
\beq
E_{N_S} = \sum_{i\alpha} \frac{v \phi_i}{\nu_\alpha - v_i}+ 
\mu_B B (2 N_S-\Omega_S) 
-\frac{1}{4} \sum_{ij} J_{ij} - \frac{1}{2} \sum_i J^{ }_{ii}
\eeq
instead of (\ref{nonunispineigen1st}), 
with arbitrary parameters $\phi_i$. The couplings 
are given by Eq.(\ref{spineigen1st}),  
replacing $\zeta_i$ with $\phi_i$.
 
\subsection{The charge Hamiltonian}\label{chargesu2}

To construct an integrable model for non uniform couplings 
we generalize formula (\ref{bcs}) to: 
\begin{equation}
\t{H}_K=\sum_i 2\eta_i 
\t{\tau}_i + \sum_{ij} A_{ij} 
\t{\tau}_i \t{\tau}_j -\sum_i \frac{3}{4} \phi_i ,
\label{nonuniA}
\end{equation} 
with real symmetric $A_{ij}$, wherein 
the operators $\t{\tau}_i$ are obtained by substituting 
the isotropic Gaudin models $\Xi_i$ in formula (\ref{isointmot})  
by the {\em anisotropic} Gaudin Hamiltonians $\t{\Xi}_j (\fat{\sigma})$ (see also appendix)
\begin{equation}
\t{\Xi}_j (\fat{\sigma}) = {\sum_{k}}' w^a_{jk} \sigma_j^a \sigma_k^a .
\end{equation}
\begin{equation}\label{newintegrals}
\t{\tau}_i=K_i^z + \t{\Xi}_i({\bf K}) .
\end{equation}
A sufficient condition for these operators to commute with each other is that 
the coefficients $w_{ij}^\alpha$ of $\t{\Xi}_j (\fat{\sigma})$
can be parametrized as 
$w_{ij}^z\equiv v_{ij}= p u \coth{\left[p(u_i-u_j)\right]}$, 
$w_{ij}^x = w_{ij}^y \equiv w_{ij}= p u /\sinh\left[p(u_i-u_j)\right]$. 
While $\{u_i\}$, $u$ are real, the parameter $p$ can be real or imaginary
and switches between hyperbolic and trigonometric functions respectively\cite{NOTE}. 
The resulting Hamiltonian is
\beqa
\t{H}_K &=& \sum_i 2 \eta_i K_i^z - \frac{1}{2}
\sum_{ij} g_{ij} (K^+_i K^-_j + K_i^- K_j^+) + 
4 \sum_{ij} U_{ij} K_i^z K_j^z \label{gen-charge-Ham}\\
&&+ H_3 + H_4 
\eeqa
with the couplings 
\beq\label{coupling-constants}
\begin{array}{rcl}
g_{ij} &=&  - \bigfrac{p u (\eta_{i}-\eta_{j})}{
	\sinh[p (u_{i}-u_{j})]} \;, \\
&& \\ 
4 U_{ij} &=&  A_{ij} + p u (\eta_{i} - \eta_{j}) 
\coth[p (u_{i}-u_{j})] \; ,
\end{array}
\quad \mbox{for }i\neq j 
\eeq

The Hamiltonians $H_3$ and $H_4$ contain interactions up to tri- and 
quadri-linear in the charge $su(2)$ operators: 
\beqa
H_3 &=& \dsum_{ij} A_{ij} (K_i^z \t{\Xi}_j({\bf K})+
\t{\Xi}_i({\bf K}) K_j^z)
\nonumber \\
&=&  {\dsum_{i j k}}' \left[ A_{ij}-A_{ik}\right] \left(  
	 v_{jk} K_i^z K_j^+ K_k^- +  w_{jk} K_i^z K_j^z K_k^z \right) \label{H3} \\
&+&  \frac{1}{2} \; 
	{\dsum_{ij}}' \left[ A_{ij}-A_{jj}\right]  w_{ij}  K_i^z  \nonumber,
\eeqa
\beqa
\nonumber
H_4 &=& \dsum_{ij} A_{ij} \t{\Xi}_i({\bf K}) \t{\Xi}_j({\bf K}) 
 \\
&=&   {\dsum_{ijkl}}'  \frac{1}{2} 
	\left[ A_{ik} - A_{il} + A_{jl} - A_{jk}\right] v_{ij}w_{kl}
	K_i^z K_j^z K_k^+ K_l^- \nonumber \\
&+& 
	\frac{1}{4} \; {\dsum_{ij}}'\left\{ 2A_{ij}-A_{ii}-A_{jj}\right\} 
	w_{ij}v_{ij} K_i^+ K_i^- \label{H4} \\
&+& \frac{1}{4} \; {\dsum_{ij}}'{\dsum_{kl}}'
	\left\{ A_{ik}-A_{jk}-A_{il}+A_{kl}\right\} 
	w_{ij}w_{kl} K_i^+ K_j^- K_k^+ K_l^- \nonumber \\
&+& \frac{1}{4} \; {\dsum_{ij}}'{\dsum_{kl}}' 
	\left\{ A_{ik}-A_{jk}-A_{il}+A_{kl}\right\} 
	v_{ij}v_{kl} K_i^z K_j^z K_k^z K_l^z. \nonumber
\eeqa
It is interesting to note that $H_3$ and $H_4$ vanish for $A_{ij}= A$,
which can be also seen directly from equations (\ref{H3}) and
(\ref{H4}), because of $\sum_i \Xi_i = 0$ (see appendix). 
Another possible simplification can be achieved for 
$A_{ij}= \alpha_i + \alpha_j$, for which  
the overall effect of the second term in Eq.(\ref{nonuniA}) is to 
renormalize the parameters $\eta_j$:
\beq
\t{H}_K = \sum_j 2\left(\eta_j + K^z  \alpha_j\right) \t{\tau}_j .
\eeq
We point out that the Coulomb term 
can be canceled choosing 
$A_{ij}=g_{ij} \cosh{[p(u_i - u_j)]}$~\cite{NOTE-NEUTRAL}.
In ${\mathcal H}_K$, the eigenstates of $\t{\tau}_i({\bf K})$, and 
hence of $\t{H}_K$, are 
\beq
\ket{\Psi_{N_K}} = \prod_{\alpha}\sum_{j} 
\frac{p~K_j^+}{\sinh{[p(\omega_\alpha - u_j)]}} \ket{0}_K .
\label{gen-vector}
\eeq
The eigenvalues $\t{t}_i$ of $\t{\tau}_i({\bf K})$ are given by 
\beq
\t{t}_i =  \frac{u}{2} \sum_{\alpha} p 
	\coth{[p ( \omega_\alpha - u_i )]} 
	- \frac{u}{4} {\sum_j}' p \coth{[p ( u_j - u_i )]} - \frac{1}{2} . 
\label{gen-eigen}
\eeq
The quantities $\omega_\alpha$ must fulfill the equations
\beq
\frac{2}{u} + \sum_j p\;\coth{p[(u_j-\omega_\alpha)]} 
= 2{\sum_{\beta}}' 
p\; \coth{[p ( \omega_\beta - \omega_\alpha )]} \;. 
\label{generalized-richardson}
\eeq
We discuss the case of constant $A_{ij}\equiv A$. 
Then, the Hamiltonian simplifies 
and formula~(\ref{nonuniA}) becomes
\beqa
\t{H}_K &=& \sum_i 2\eta_i \; \t{\tau}_i 
+ A \sum_{ij} \t{\tau}_i\; \t{\tau}_j 
= \sum_i 2\eta_i \; \t{\tau}_i
+ A \left(\sum_i \t{\tau}_i\right)^2 \nonumber\\
&=& \sum_i 2\eta_i \; \t{\tau}_i  + A (K^z)^2.
\eeqa
In this case $\t{H}_K$ is the Hamiltonian (\ref{chargeh}), provided 
the coupling constants in the latter 
are parametrized as in Eq.~(\ref{coupling-constants}) 
for $A_{ij}\equiv A$.
The eigenstates are still given by (\ref{gen-vector}) with the eigenvalues 
\begin{eqnarray}
E_{N_K} &=& p u \sum_{j\alpha} \eta_{j} \coth{[p ( \omega_\alpha - u_j )]}
- A N_K ( \Omega_K - N_K) - E_K.
\label{eigencharge}
\end{eqnarray}
In the $p\to 0$ limit, which corresponds to the isotropic limit 
for the Gaudin Hamiltonians, the coupling constants become
\beq
\begin{array}{rcl}
g_{ij} &=& {- u \left(\eta_{i}-\eta_{j}\right)}/(u_{i}-u_{j}) \;, \\
&& \\
4 U_{ij} &=& A +  u (\eta_{i} - \eta_{j})/ (u_{i}-u_{j}),
\end{array} 
\; \mbox{ for } i\neq j \label{isononunicoupl}
\eeq
The eigenstates and eigenvalues are 
\begin{eqnarray}
\ket{\Psi_{N_K}}&=&\prod_\alpha \sum_i 
\frac{K_i^+}{\omega_\alpha - u_i} \ket{0}_K \nonumber \\
E_{N_K}&=&\sum_{i\alpha} \frac{\eta_i}{\omega_\alpha - u_i} - A N_K (\Omega_K - N_K)-E_0 , 
\label{eigeniso}
\end{eqnarray}
where $\omega_\alpha$ have to be  solutions of the equations
\begin{equation} 
\frac{2}{u} + \sum_{j} \frac{1}{u_j - \omega_\alpha} 
= 2 {\sum_{\beta}}' \frac{1}{\omega_\beta - \omega_\alpha} . 
\label{quasire}
\end{equation}
Uniform couplings $g_{ij}=g$, $U_{ij}=E_{c}$ are then obtained 
from a linear $u$--$\eta$ relation. Choosing $u_j=-2\eta_j/g$, 
$\omega_\alpha= - E_\alpha/g$, and $u=2$, Eqs.(\ref{quasire}) 
are the RS equations (\ref{richardson}) (we recall that $eta_i=\xi_i + const.$). 
Setting $A=4E_{c}+g$ as well, the Hamiltonian (\ref{chargeh})
is the BCS Hamiltonian (up to a constant).

\subsection{The spin Hamiltonian}\label{spinsu2}

We remark, that the procedure leading here to the exact solution
of the spin Hamiltonian coincides with that applied to $H_K$, 
because the underlying algebraic structure is still $su(2)$.
\\
We study the Hamiltonian
\begin{equation}
H^{xxz}_S = \sum_i 2\zeta_i \tau_i - \frac{3}{4} J \Omega_S,
\label{hspin}
\end{equation} 
with $\tau_i = S_i^z + \t{\Xi}_i({\bf S})$.
Also here, the conservation of the $z$-component of the total spin
restricts the anisotropy of the Gaudin models to the XXZ type.
The Hamiltonian we get from Eq.~(\ref{hspin}) is 
\begin{equation}
H^{xxz}_S = \sum_i 2 \zeta_i S_i^z - \sum_{ij} \left( 
J_{ij} (S_i^x S_j^x + S_i^y S_j^y) + J^z_{ij} S_i^z S_j^z \right) , 
\label{gen-spin-h}
\end{equation}
where the couplings are defined by 
\beq
\left.
\begin{array}{rcl}
J_{ij} &=& - \bigfrac{ p v{\left(\zeta_{i}-\zeta_{j}\right)}}
{\sinh [p (v_{i}-v_{j})]} , \\
&& \\
J^z_{ij} &=& - p v (\zeta_{i} - \zeta_{j}) \coth{[p (v_{i}-v_{j})]},
\end{array} 
\right\} i\neq j 
\label{spin-coupling-constants}
\eeq
The eigenstates of $H^{xxz}_S$ in ${\mathcal H}_S$ and its eigenvalues are: 
\begin{eqnarray}
\ket{\Phi^{xxz}_{N_S}} &=& \prod_{\alpha}\sum_{j} 
\frac{p~S_j^+}{\sinh{[p(\nu_\alpha - v_j)]}} \ket{0}_S ,
\label{gen-spin-vector} \\
E^{xxz}_{N_S} &=& 
p v \sum_{j\alpha} \zeta_{j} \coth{[p ( \nu_\alpha - v_j )]}
-\sum_j \zeta_j
-\frac{1}{4} \sum_{ij} J^z_{ij} - \frac{1}{2} \sum_i J^{ }_{ii} 
\label{gen-spin-eigen}
\end{eqnarray}
where the $\nu_\alpha$ have to satisfy the RS equations
\begin{equation} 
\frac{2}{v} + \sum_{j} \frac{1}{v_j - \nu_\alpha} 
= 2 {\sum_{\beta}}' \frac{1}{\nu_\beta - \nu_\alpha} . 
\label{modigaudin}
\end{equation}
In order to attack the Hamiltonian (\ref{spinh}), 
i.e. having an XXX- instead of an XXZ-type coupling, we have to perform the
isotropic limit for the Gaudin Hamiltonians and obtain
$H^{xxx}_S = \sum_i 2 \zeta_i S_i^z - \sum_{ij} J_{ij}\: 
\vec{S}_i \cdot \vec{S}_j$,
with the coupling defined as
\beq
J_{ij}= - v \frac{\zeta_i - \zeta_j}{v_i - v_j}\quad,\quad i\neq j .
\label{spinterm}
\eeq
The eigenvalues and the eigenstates of $H^{xxx}_S$ are 
\begin{eqnarray}
\ket{\Phi^{xxx}_{N_S}}&=&\prod_{\alpha} \sum_j 
\frac{c^\dagger_{j\up} c^{}_{j\dwn}}{\nu_\alpha - v_j}\ket{0}_S \quad ,
\label{eigenspin} \\
E^{xxx}_{N_S} &=& 
\sum_{j\alpha} \frac{v \zeta_i}{\nu_\alpha - v_i}
-\sum_j \zeta_j -\frac{1}{4} \sum_{ij} J_{ij} - 
\frac{1}{2} \sum_i J_{ii} \quad , 
\label{nonunispineigen}
\end{eqnarray}

For uniform Zeeman splitting, 
$\zeta_i\equiv\mu_B B$ 
($\mu_B$ is Bohr's magneton, $B$ the magnetic field), the Hamiltonian can be 
expressed as 
\begin{equation}
H^{xxx}_S=2\mu_B B S^z + \sum_{j} (2 \phi_j \Xi_j(\fat{S}) - 3 J_{jj}/4) . 
\end{equation} 
We remark that the constants of motions are Gaudin Hamiltonians themselves 
(instead of the $\tau$'s). 
The resulting couplings are 
given by Eq.~(\ref{spinterm}) with $\zeta_i=\phi_i$. 
The eigenvalues of $H_S$ are then
\begin{equation}
E_{N_S} = \mu_B B (2 N_S-\Omega_S) 
+ \sum_{i\alpha} \frac{ \phi_i}{\nu_\alpha - v_i} 
- \frac{1}{4}\sum_{ij} J_{ij} -\frac{1}{2} \sum_i J_{ii} \quad , 
\label{spineigen}
\end{equation}
where $\nu_\alpha$ have to be solutions of 
\begin{equation}	
\sum_{j} \frac{1}{v_j - \nu_\alpha} 
= 2 {\sum_{\beta}}' \frac{1}{\nu_\beta - \nu_\alpha} \quad . 
\label{gaudineqs}
\end{equation}
The corresponding eigenstates are given by Eq.(\ref{eigenspin}). 
The main difference is that the
$\nu_\alpha$ fulfill the Gaudin Eqs.(\ref{gaudineqs}) 
instead of the RS  Eqs.(\ref{modigaudin}).

\section{Discussion of the coupling constants}\label{discuss}

In the following we focus on possible choices of the coupling constants 
in $H_K$ that seem appropriate for the physics of 
metallic grains\cite{spincouplgs}. 
It is physically reasonable to demand that the  
pairing attraction decays with 
the energy difference of the interacting levels. 
This can be achieved by the choice $p=1$ in Eq.(\ref{couplings}) --- which
corresponds to have hyperbolic functions in the couplings --- and 
$u_i$ being a monotonic function of the single particle energies. 
In order to make the couplings $g_{ij}$ and $U_{ij}$ depend on the 
energy difference $\eps_i-\eps_j$ alone, we choose 
$u_j=-\eps_j/\delta \epsilon$; 
$\delta \epsilon$ has the dimension of an energy ($u_i$ are dimensionless), 
and it turns out to be a measure of the range of the pairing interaction. 
\begin{figure}
\begin{center}\subfigure[]{
\includegraphics[width=1.5in,angle=-90]{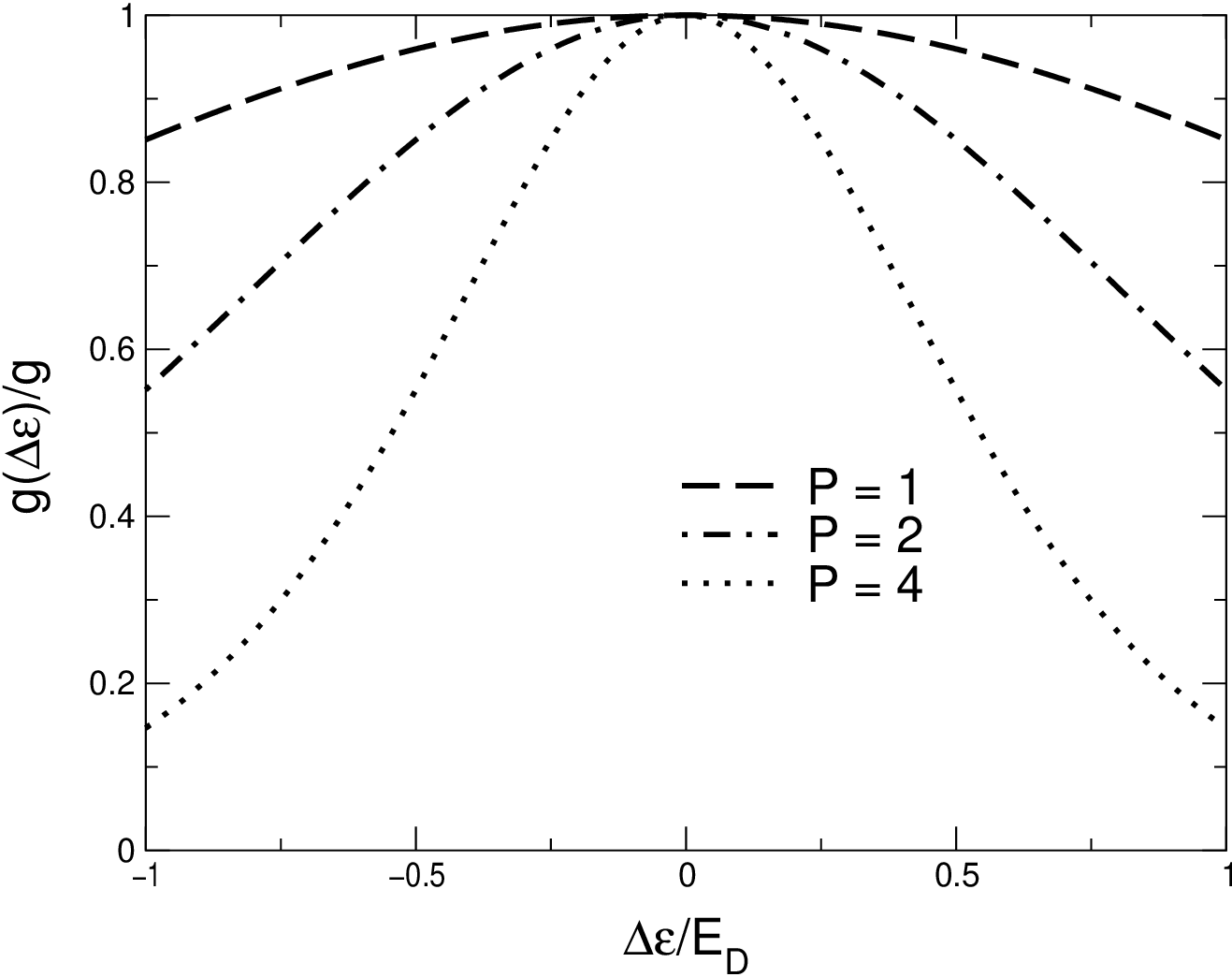}}
\hspace{1in}
\subfigure[]{
\includegraphics[width=1.5in,angle=-90]{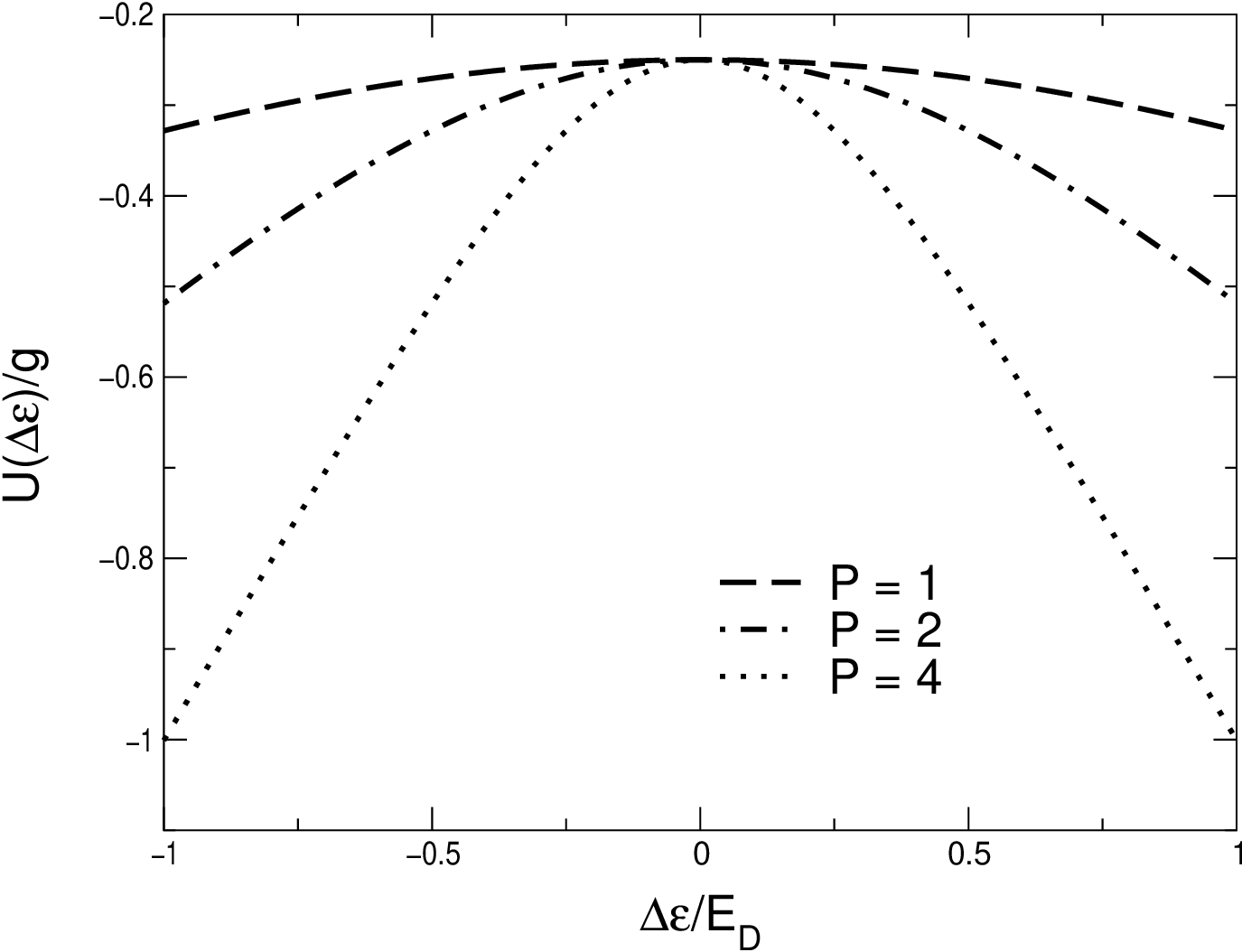}}
\end{center}
\caption{\label{nonuniform-single} (a) The pairing  and (b) the Coulomb  
interaction as function of the energy difference $\Delta \varepsilon$ of the 
single-particle levels. 
In Figure (b) the $y$ axis is offset by $A/4g$
 (see the definition of $U$ in Eqs.~(\ref{couplings}))}
\end{figure}
We introduce the dimensionless parameter $P$ such that 
 $\delta \epsilon = E_D/P$, $2E_D$ being the width $\Delta\eps_{max}$
of the single particle spectrum. 
We choose $u = P g / E_D$, in order that the limit $P\longrightarrow 0$
corresponds to $g_{ij}\equiv g$. 
The resulting pairing is plotted in Fig.\ref{nonuniform-single}(a). 
The Coulomb repulsion slowly decreases with the energy difference, 
as shown in Fig.\ref{nonuniform-single}(b). For large energy difference  
and $P\geq 1$ the decay is linear in the single particle energy difference. 

We now discuss the interpretation of Hamiltonian ~(\ref{ourH})
as a model for ${\mathcal N}$ weakly coupled grains. 
We introduce the grain index $a=1,\dots,{\mathcal N}$ (the index 
increasing  with the spatial distance of the grains) 
and the sets $I_a$ such that for $i\in I_a$, 
$\eps_i$ is an energy level of grain $a$;
$\bigcup_{a=1}^{\mathcal N} I_a =I$.
The charge part e.g. of Hamiltonian (\ref{ourH}) can then be rewritten as
\beqa
H_{\mathcal N} &=& \sum_{a=1}^{\mathcal N} \sum_{i_a} 
\xi^{(a)}_{i_a} c_{a,i_a \sigma}^\dagger c_{a,i_a \sigma} 
- \sum_{a,b=1}^{\mathcal N}\sum_{i_a,j_b}   
g^{(a,b)}_{i_a j_b}\, 
c_{a,i_a\up}^\dagger c_{a, i_a \dwn}^\dagger 
c_{b,j_b \dwn} c_{b,j_b \up} \nonumber \\
&& +\sum_{a,b=1}^{\mathcal N}\sum_{i_a,j_b}
U^{(a,b)}_{i_aj_b} 
n_{a,i_a \sigma} n_{b,j_b \sigma'} .\label{multi-grain}
\eeqa
This Hamiltonian describes a system of many grains,
where $i_{a}=1,\dots\Omega_a$ labels the elements of $I_{a}$ 
and $c_{a,i_a \sigma}$ annihilates an electron with spin $\sigma$ in 
the $i_a$-th level of $a$-th grain, $\xi^{(a)}_{i_a}$. 
For $a\neq b$, $g^{(a,b)}$ describes the tunneling of Cooper pairs;
$g^{(a,b)}_{i_a j_b} = g_{ij}$ if $i$ is the $i_a$-th element of $I_a$,
and $j$ the $j_b$-th element of $I_b$; 
$U^{(a,b)}$ is the Coulomb repulsion between the grains $a$ and $b$.
It is written in terms of $U_{ij}$ analogously to $g^{(a,b)}$.
Couplings $g^{(a,a)}$ and $U^{(a,a)}$ describe intra-grain 
pairing and Coulomb interactions, respectively. 
Fixing the parameters as 
$u_{j} = \Phi_{a} - P \varepsilon_{j}/E_{D}$ for $j\in I_a$
and imposing
$\Phi_{a+1}-\Phi_{a} \gg P$, 
the tunneling amplitude of pairs is exponentially suppressed 
compared to the intra-grain pairing coupling, as seen in Fig. 2(a). 
\begin{figure}
\begin{center}\subfigure[]{
\includegraphics[width=1.5in,angle=-90]{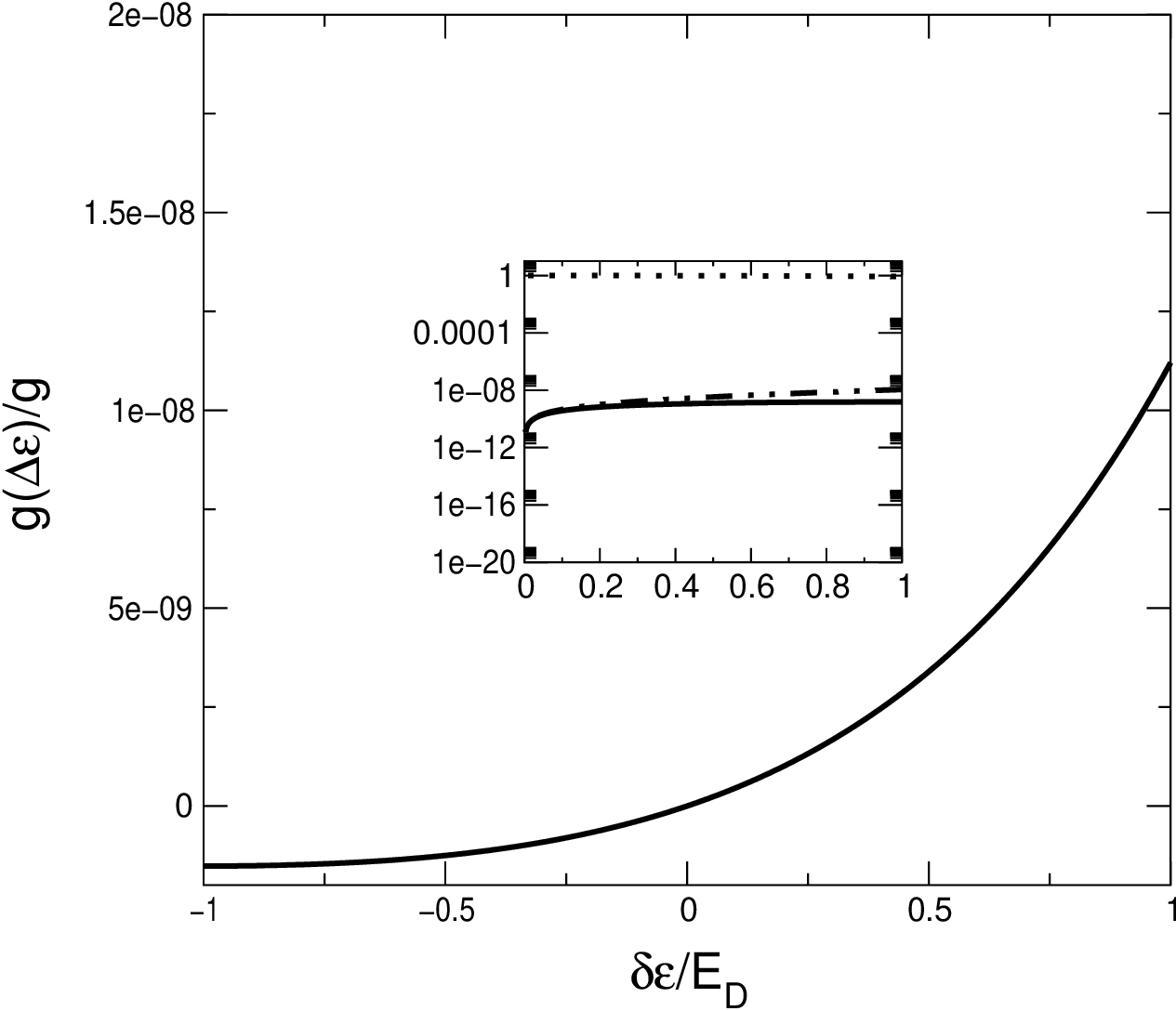}}
\hspace{1in}
\subfigure[]{
\includegraphics[width=1.5in,angle=-90]{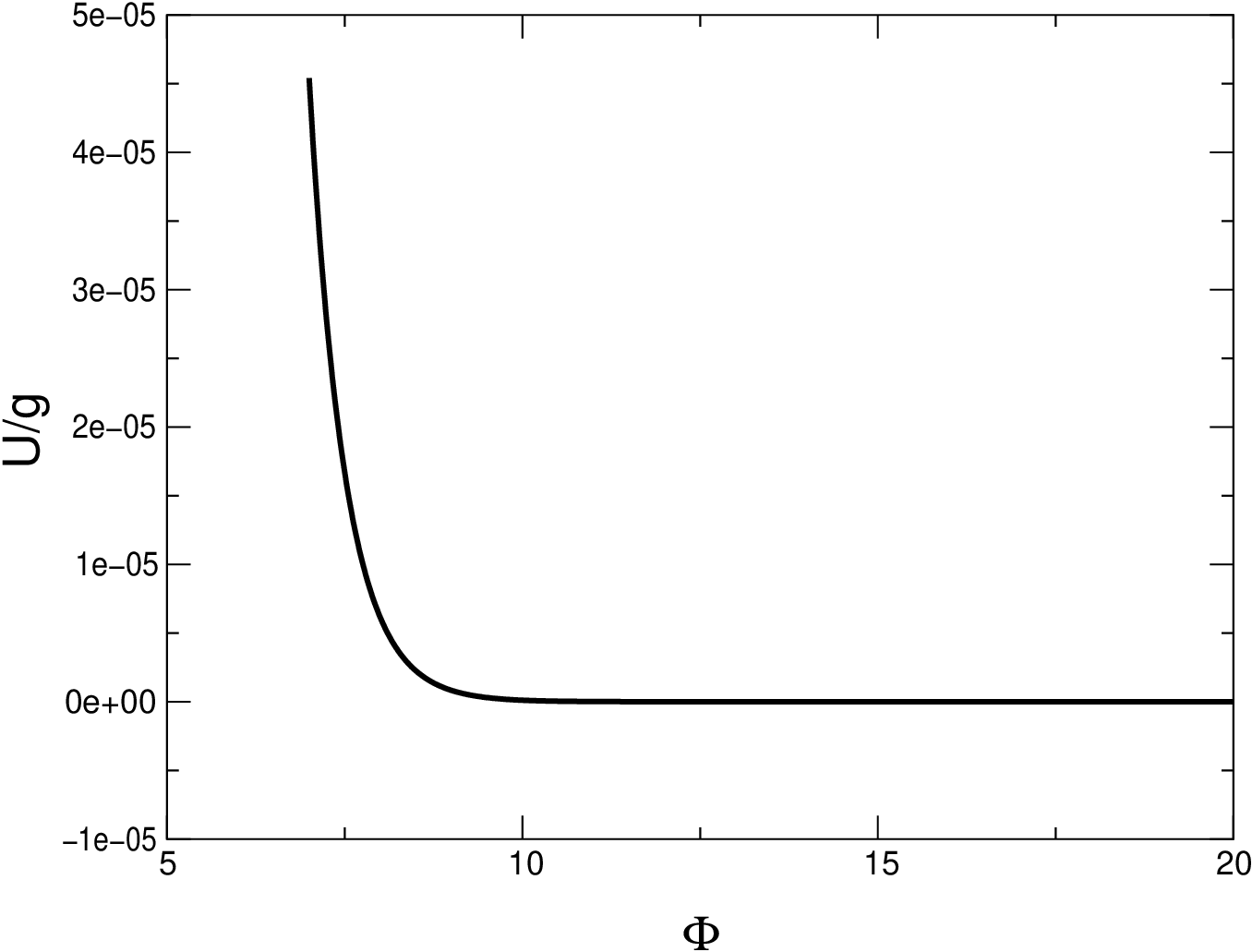}}
\end{center}
 \caption{\label{interpair} (a) The inter-grain pairing coupling $g^{(1,2)}$,
i.e. the pair-tunneling amplitude,
between two grains with $P=1$ and $\Phi=\Phi_2 - \Phi_1 = 20$. 
The inset shows that the pair tunneling amplitude 
$g^{(1,2)}$ is much smaller than $g^{(1,1)}\equiv g^{(2,2)}$. 
(b) The intra-grain Coulomb repulsion  
is nearly constant for $\Phi \gg P=1$ for fixed pairs of interacting 
levels; this constant increases linearly with $\Delta \varepsilon$ 
(the $y$ axis is offset by $A/4g$).}
\end{figure} 

\section{Conclusions}\label{concls}

In this paper, we have constructed a new class of integrable models
as a functional of mutually commuting operators, 
which are a generalization of the BCS integrals of motion 
found in Ref.~\cite{CRS}. The latter 
are tightly connected with isotropic Gaudin Hamiltonians and
the generalization presented in this paper consists in  
replacing isotropic with anisotropic Gaudin Hamiltonians.
The exact solution of the Hamiltonian constructed in this way
has been obtained via the diagonalization of these integrals of motion.
This procedure is applied to two orthogonal 
realizations of $su(2)$ (see section \ref{model}), 
leading to the model Hamiltonians $H_K$ and $H_S$ --
Eqs.~(\ref{genHamK}) and (\ref{genHamS}).
The resulting models describe systems with nonuniform interactions,
whose relevance was discussed in Ref.~\cite{SIVAN} 
according to recent experiments on metallic grains~\cite{RBT}.
The nonuniform couplings we obtain are tunable
by a set of parameters, which can be chosen such that 
physical demands on the couplings (see Ref.~\cite{GLAZMAN,ALTSHULER}
and section \ref{discuss}) are fulfilled. This is demonstrated
in Figs.~\ref{nonuniform-single} and  \ref{interpair} 
where the couplings are plotted for an exemplary choice
of the parameters in the hyperbolic regime, $p=1$. 
\\
We also discuss possible applications of this class of models
to systems of few coupled grains and present a choice of 
parameters such that the intra-grain couplings significantly
dominate the inter-grain couplings.

It is noteworthy that the class of Hamiltonians (\ref{genHamK}) and 
(\ref{genHamS}) can also describe a spin chain with a long-range, 
nonuniform XXZ interaction. Then, $\eta_j$ and $\zeta_j$ 
play the role of a nonuniform magnetic field and the
indices ($i$ and $j$) are to be interpreted as 
labeling {\em lattice sites} rather than energy levels.
In this case the use of trigonometric functions 
($p=\i$ in Eqs.~(\ref{couplings})) can account for boundary conditions.

We want to emphasize that also in 
nuclear physics~\cite{IACHELLO,RICHARDSON-PRIVATE}, 
QCD~\cite{QCD}, and astrophysics\cite{ASTROPHYSICS},
the BCS Hamiltonian (with uniform couplings) 
has been thoroughly investigated.
The three and four-body interactions, Eq.(\ref{H3}) and (\ref{H4}), 
might come in handy in these contexts. 

Further generalizations of the present work seem interesting.
Studies on the completely anisotropic 
XYZ Gaudin model can lead  to Hamiltonians where 
the total number of particles (or the total $z$ component 
of the spin) is not conserved. These models are 
connected with the $2D$ eight-vertex model~\cite{GAUDIN-BOOK,SKLYANIN-XYZ},
and eventually there is a connection with its off-shell 
Bethe Ansatz alike that between the six-vertex model
and the uniform BCS model~\cite{AMICO,BABUJIAN}. 
\\
In order to describe interacting bosons, 
$su(1,1)$ Gaudin models instead of  $su(2)$ ones
must be considered; for the  isotropic $su(1,1)$ Gaudin model see 
Ref.~\cite{DUKELSKY}. 
Consequently, models based on anisotropic $su(1,1)$ Gaudin models 
should be studied as well.
\\
Finally correlation functions for the present class of models can be studied 
along the lines described in Ref.\cite{CORRELATION-BCS}.

\thanks{
We acknowledge  G. Falci and R. Fazio
for constant support and  invaluable help.
We acknowledge R. W. Richardson for constructive discussions. 
We furthermore would like to thank D. Boese,
and A. Mastellone for useful discussion.
}


\appendix
\section{Gaudin models} \label{b-pendix}

Here we summarize the main results of Ref.\cite{GAUDIN}, which are relevant
for this work.
\\
Gaudin studied the following class of Hamiltonians:
\begin{equation}
\t{\Xi}_j (\fat{\sigma}) = {\sum_{k}}' w^a_{jk} \sigma_j^a \sigma_k^a \quad ,
\label{anisogaudin}
\end{equation}
where $\sigma^\alpha$ are the Pauli matrices, and the 
sum over $a=x,y,z$ is implied. Mutual commutation of $\t{\Xi}_j$ requires  
\begin{equation}
w_{ij}^a w_{jk}^c + w_{ji}^b w_{ik}^c = w_{ik}^a w_{jk}^b .
\label{conds}
\end{equation} 
Identifying $(x,y,z)$ with $(1,2,3)$, let us define 
\begin{equation}
w_{ij}^a =: \frac{\Theta^{a+1}(u_{ij})}{\Theta^{a+1}(0)\Theta^{1}(u_{ij})} , 
\end{equation}
with $u_{ij} = - u_{ji}$; $\Theta^{1}(u)$ is an odd function, 
and $\Theta^{a+1}(u)$ are three even functions.
For $u_{ij}$ such that $u_{ij} + u_{jk} + u_{ki}=0$ 
(e.g. $u_{ij}=u_i - u_j$), it is found 
that Eqs.(\ref{conds}) are satisfied if $\Theta^a$ are the elliptic 
Jacobi functions. It is possible to find eigenstates and eigenvalues 
of $\t{\Xi}_i$ through Bethe Ansatz\cite{GAUDIN,SKLYANIN-GAUDIN}. 
\\
Imposing conservation of the $z$-projection of the total spin,
$\comm{\sum_i\sigma_i^z}{\t{\Xi}_j}=0$, 
we get the conditions $w_{jk}^1=w_{jk}^2$, i.e. the anisotropy
is of XXZ-type.
In this case, the (rescaled) elliptic functions reduce to the 
trigonometric ones
\begin{eqnarray}
w_{ij}=&w_{ij}^1&=w_{ij}^2= p w /{\sinh}[p(u_i-u_j)] \nonumber \\
v_{ij}=&w_{ij}^3&= p w \coth{[p(u_i-u_j)]}. 
\end{eqnarray}
Here, we included a real scale factor $w$ together with 
the parameter $p$ which can either be real or imaginary.
Defining
\begin{equation}
\t{\sigma}^{\pm}(u)=\sum_i \frac{p w}{\sinh [p (u_i - u)]} 
\sigma_i^\pm , \quad 
\t{\sigma}^{z}(u)=\sum_i p w \coth{[p (u_i - u)]} \sigma_i^z, 
\label{anisocreat}
\end{equation}
the common eigenstates $\ket{\Psi_{N_\up}}$ of $\t{\Xi}_i$ 
for $N_\up$ up spins read 
\beq
\ket{\Psi_{N_\up}} = \prod_{\alpha} \t{\sigma}^+(\omega_\alpha)\ket{0}, 
\label{gaudineigenfunction} 
\eeq
where $\ket{0}$ is the lowest weight state.  
The corresponding eigenvalues are 
\beq
\t{\xi}_i = {\sum_j}' pw \coth{[p(u_i - u_j)]} - 
	2 \sum_\alpha pw \coth{[p(u_i - \omega_\alpha)]},
\label{gaudinanisoeigenv} 
\eeq
where $\omega_\alpha$ must satisfy 
\beq
\sum_j p\;\coth{p[(u_j-\omega_\alpha)]} =
	2{\sum_{\beta}}' p\; \coth{[p ( \omega_\beta - \omega_\alpha )]};
\label{gaudinanisoeqs}
\eeq
$\alpha$, $\beta$ run from $1$ to $N_\up$. 
The isotropic case is obtained imposing $\comm{\fat{S}}{\Xi_i}=0$, and 
corresponds to the limit $p\to 0$. It leads to the isotropic
Gaudin Hamiltonians
\begin{equation}
\Xi_j(\fat{\sigma}) = 
{\sum_{k}}' \frac{\vec{\sigma}_j \cdot \vec{\sigma}_k}{u_j-u_k} .
\label{isogaudin}
\end{equation}
Using here the operators
\begin{equation}
\sigma^{\pm}(u)=\sum_i \frac{\sigma_i^\pm}{(u_i - u)} , \quad 
\sigma^{z}(u)=\sum_i \frac{\sigma_i^\pm}{(u_i - u)}, 
\label{isocreat}
\end{equation}
the common eigenstates of ${\Xi}_i$ for $N_\up$ up spins can be written as: 
\beq
\ket{\Psi_{N_\up}} = \prod_{\alpha} \sigma^+(\omega_\alpha)\ket{0},
\label{isogaudineigenfunction} 
\eeq
and the corresponding eigenvalues are: 
\beq
\xi_i = {\sum_j}' \frac{1}{u_i - u_j} - 2 \sum_\alpha \frac{1}{u_i - \omega_\alpha},
\label{gaudinisoeigenv} 
\eeq
where $\omega_\alpha$ have to satisfy
\beq
\sum_j \frac{1}{u_j-\omega_\alpha} = 
	2{\sum_{\beta}}' \frac{1}{\omega_\beta - \omega_\alpha}
\label{gaudinisoeqs}
\eeq
We note, that the specialized Gaudin equations (\ref{gaudinisoeqs}) and their more general
version (\ref{gaudinanisoeqs}) are the $g\to \infty$ limit of the 
RS equations (\ref{richardson}) and their generalization 
(\ref{generalized-richardson}), respectively.



\begin{thebibliography}{99}

\bibitem{RBT} C.T. Black, D.C. Ralph, and M. Tinkham, 
	Phys. Rev. Lett. {\bf 74}, 3291 (1995); {\bf 76}, 688 (1996);
	{\bf 78}, 4087 (1997). 
\bibitem{REPORT} J. von Delft and D.C. Ralph, Phys. Rep. {\bf 345}, 61 (2001).
\bibitem{DAVIDOVIC} D.G. Salinas, S. Gu\'eron, D.C. Ralph, 
	C.T. Black, and M. Tinkham, Phys. Rev. B {\bf 60}, 6137 (1999);
	D. Davidovic and M. Tinkham, Phys. Rev. Lett. 
	{\bf 83}, 1644 (1999). 
\bibitem{GUERON} S. Gu\'eron, M.M. Deshmukh, E.B. Myers, and 
	D.C. Ralph, Phys. Rev. Lett. {\bf 72}, 386 1998.
\bibitem{VONDELFT} A. Di Lorenzo, R. Fazio, F. Hekking, G. Falci, 
	and A. Mastellone, Phys. Rev. Lett. {\bf 84}, 550 (2000);
	M. Schechter, Y. Imry, Y. Levinson, and J. von Delft 
	Phys. Rev. B {\bf 63}, 214518 (2001). 
\bibitem{GLAZMAN} L. I. Glazman in Ref.~\cite{PROCGRAINS}.
\bibitem{ALTSHULER} I.L. Kurland, I.L. Aleiner, and B.L. Altshuler, 
	Phys. Rev. B {\bf 62}, 14886 (2000).
\bibitem{STONE} P. Jacquod, A. Douglas Stone, {\it cond-mat/0102100}. 
\bibitem{ALHASSID}Y. Alhassid, Rev. Mod. Phys. {\bf 72}, 895 (2000).
\bibitem{PROCGRAINS} {\it Proceedings of the International 
Conference on Electron Transport in Mesoscopic Systems, ETMS'99}
 ~J. Low Temp. Phys. {\bf 118} (2000).
\bibitem{RICHARDSON} R.W. Richardson and N. Sherman, Nucl. Phys. 
	{\bf 52}, 221 (1964); {\bf 52}, 253 (1964).
\bibitem{CRS} M.C. Cambiaggio, A.M.F. Rivas, and M. Saraceno,
	Nucl. Phys. A {\bf 624}, 157 (1997).
\bibitem{ADO} L. Amico, A. De Lorenzo, and A. Osterloh,  Phys. Rev. Lett. 
	{\bf 86}, 5759 (2001).
\bibitem{AMICO} L. Amico, G. Falci, and R. Fazio, J. Phys. A {\bf 34},  6425 (2001).
\bibitem{SIERRADUKELSKY} 
	G. Sierra, J. Dukelsky, G.G. Dussel, J. von Delft, and 
	F. Braun, Phys. Rev. B {\bf 18}, R11890 (2000).
\bibitem{AGAM} O. Agam, N.S. Wingreen, B.L. Altshuler, D.C. Ralph, and 
	M. Tinkham, Phys. Rev. Lett. {\bf 78}, 1956 (1997); 
	B.L. Altshuler, Y. Gefeln, A. Kamenev, and L.S. Levitov, 
	Phys. Rev. Lett.
	{\bf 78}, 2803 (1997); Ya.M. Blanter, Phys. Rev. B {\bf 54}, 12807 
	(1996). 
\bibitem{TRIVIALEXTENSION} The only extension we found in this way, was
	a simply ``rotated'' model, where each eigenstate was rotated
	by a phase. In this case, the (complex) pairing coupling had the
	simple form $g_{jk}=g \exp \i (\varphi_j-\varphi_k)$. The RS 
	equations are consequently unchanged.
\bibitem{BRAUN} F. Braun and J. von Delft, Phys. Rev. Lett. {\bf 81}, 
	4712 (1998).
\bibitem{GAUDIN} M. Gaudin, J. Phys. {\bf 37}, 1087 (1976).
\bibitem{SKLYANIN-GAUDIN} E.K. Sklyanin, J. Sov. Math. {\bf 47}, 2473 (1989).
\bibitem{NOTE} It was not included in $u, \{u_j\}$ in order to make clear the relation 
	with the isotropic case, obtained in the limit $p\to 0$. 
\bibitem{NOTE-NEUTRAL} In this case, the contributions from Hamiltonians 
	$H_3$ and $H_4$ do not vanish. Nevertheless, these terms 
	might be useful to describe a system of neutral particles 
	with three and four-body interactions. 
\bibitem{spincouplgs} The possible range of couplings is very large:
	there are as many free parameters as single-particle
	levels. In $H_S$ for constant Zeeman splitting, the range
	is even wider, because the $\phi_j$ are an additional 
	set of independent parameters. We do not show an 
	exemplary plot of the spin couplings.
\bibitem{SIVAN} U. Sivan, F.P. Milliken, K. Milkove, 
	S. Rishton, Y. Lee, J.M. Hong, V. Boegli, D. Kern, and M. deFranza,
	Europhys. Lett. {\bf 25}, 605 (1994), O. Agam, in 
	\emph{Supersymmetry and Trace Formulae: Chaos and Disorder}, 
	I. V. Lerner, J. P. Keating and D. E. Khmelnitskyi editors, Plenum Press 1999..
\bibitem{GAUDIN-BOOK} M. Gaudin, {\it La fonction d'onde de Bethe}, 
	(Masson, Paris 1982). 
\bibitem{SKLYANIN-XYZ} E.K. Sklyanin and T. Takebe, Phys. Lett. A {\bf 219}, 
	217 (1996). 
\bibitem{BABUJIAN} H.M. Babujian, J. Phys. A  {\bf 26}, 6981 (1993); 
	H.M. Babujian and R. Flume, Mod. Phys. Lett. {\bf 9}, 2029 (1994). 
\bibitem{DUKELSKY} J. Dukelsky and P. Schuck 
	Phys.Rev.Lett. {\bf 86}, 4207 (2001).  

\bibitem{IACHELLO}
	F. Iachello,
	Nucl. Phys. A {\bf 570}, 145c (1994).
\bibitem{RICHARDSON-PRIVATE} R.W. Richardson, {\it private communication}.
\bibitem{QCD} D. H. Rischke and R. D. Pisarski, proceedings of the 
	"Fifth Workshop on QCD", Villefranche, nucl-th/0004016.
\bibitem{ASTROPHYSICS}  H. Heiselberg and M. Hjorth-Jensen, 
	Phys. Rep. {\bf 328}, 237 (2000).
\bibitem{CORRELATION-BCS}L. Amico and A. Osterloh, submitted to Phys. 
	Rev. Lett.,  {\it cond-mat/0105141}.
%

\end{thebibliography}
\end{document}